\input harvmac.tex
%
\def\*{\star}
\def\({\left(}          
\def\){\right)}         
\def\[{\left[}          
\def\]{\right]}         
\def\BL{\Bigr(}
\def\BR{\Bigr)}

\def\tl{\tilde}
\def\wtl{\widetilde}
\def\frac#1#2{{#1 \over #2}}            
\def\half{{1 \over 2}}
\def\d{\partial}

\def\dd#1#2{{\partial #1 \over \partial #2}}

\def\th{\theta}
\def\vth{\vartheta}         
\def\ga{\gamma}         
\def\b{\beta}           
\def\al{\alpha}
\def\ep{\epsilon}
\def\la{\lambda}        
\def\de{\delta}         
\def\om{\omega}         
        
\def\vphi{\varphi} 
\def\ka{\kappa}

\def\Ga{\Gamma}

\def\De{\Delta}
\def\Om{\Omega}

\Title{\vbox{\baselineskip12pt\hbox{RU 96-110}
                              \hbox{hep-th/9612040}}}
{\vbox{\centerline
{$\Phi^{(2)}$   Perturbations of WZW Model}}}
\vskip15pt
\centerline{Vadim A. Brazhnikov \footnote{$^*$}{e-mail address:
vadim@physics.rutgers.edu }}
\centerline{Department of Physics and Astronomy,}
\centerline{Rutgers University,}
\centerline{NJ 08855-0849, USA }
\bigskip\bigskip\bigskip
\centerline{{\bf Abstract}}
We study $su(2)_k$ WZW model perturbed by a multiplet of  primary
fields. The theory has a rich variety of particles. Presence of
nontrivial decay processes is a peculiarity of the model. We prove
integrability by explicit construction of quantum conserved currents. 
The scattering theory is briefly discussed.
\bigskip
\Date{November, 1996}
\vfill
\eject

\lref\wz{Zamolodchikov,A.B.: Infinite additional symmetries in
two-dimensional conformal quantum field theory. Theor.Math.Phys.{\bf
65}, 1205 (1985)}

\lref\parzf{Fateev,V.A., Zamolodchikov,A.B.: Parafermionic currents in the two-dimensional conformal
quantum field theory and self-dual critical points in $Z_n$ invariant
statistical systems. Sov.Phys.JETP {\bf 62},215 (1985)}

\lref\lf{Fateev,V.A., Lukyanov,S.L.: Additional symmetries and exactly soluble models in
two-dimensional conformal field theory. Chur, Switzerland: Harwood
(1990) 117 p. (Soviet Scientific Reviews A, Physics: 15.2).} 

\lref\lw{Lepowsky,J., Wilson,R.L.: Adv.Math. {\bf 45}, 21 (1982)}

\lref\zaza{Zamolodchikov,A.B.:Higher orders integrals of motion in
two-dimensional models of the field theory with a broken conformal
symmetry. JETP Lett.{\bf 46}, 160 (1987)}

\lref\kul{Kulish,P.P.: Factorization of the classical and quantum S
matrix and conservation laws.  Theor.Math.Phys.{\bf 26}, 132 (1976)}

\lref\kar{Karowski,M., Thun,H.J., Truong,T.T., Weisz,P.H.: On the
uniqueness of a purely elastic S matrix in $(1+1)$ dimensions.
Phys.Lett.{\bf 67}B, 321 (1977)}

\lref\zaal{Zamolodchikov,Al.B.: Thermodynamic Bethe ansatz in
relativistic models. Scaling three state Potts and Lee-Yang
models. Nucl.Phys. {\bf 342}B, 695 (1990)}

\lref\musa{Cardy,J.L., Mussardo,G.: S matrix of the Yang-Lee edge
singularity in two dimensions. Phys.Lett. {\bf225}B, 275 (1989)}

\lref\fre{Freund,P.G.O., Klassen,T.R., Melzer,E.: S matrices for
perturbations of certain conformal field theories.  Phys.Lett. {\bf
229}B, 243 (1989)}

\lref\zhu{Sotkov,G., Zhu,Ch-J.: Bootstrap fusions and tricritical
Potts model away from criticality. Phys.Lett. {\bf 229}B, 391 (1989)}

\lref\musb{Mussardo,G.: Off-critical statistical models: factorized
scattering theories and bootstrap program. Phys.Rep. {\bf 218}, 215 (1992)}

\lref\she{Kastor,D.A., Martinec,E.J., Shenker,S.H.: RG flaw in $N=1$
discrete series. Nucl.Phys.{\bf 316}B, 590 (1989)}

\lref\lecla{LeClair,A., Nemeschansky,D., Warner,N.P.: S matrices for
perturbed $n=2$ superconformal field theory from quantum groups.
Nucl.Phys. {\bf 390B}, 653 (1993)}

\lref\leclb{LeClair,A.: restricted sine-Gordon theory and the minimal
conformal series.  Phys.Lett. {\bf 230}B, 103 (1989)}

\lref\smir{Smirnov,F.A.: Reductions of sine-Gordon model as a
perturbation of minimal models of conformal field
theory. Nucl. Phys. {\bf 337}B, 156 (1990)}

\lref\wzwa{Wess,J., Zumino,B.: Consequences of anomalous Ward
identities. Phys.Lett. {\bf 37}B, 95 (1971)}

\lref\wzwb{Witten,E.: Non-Abelian bosonization in two
dimensions. Comm.Math.Phys. {\bf 92}, 455 (1984)}

\lref\wzwc{Novikov,S.P.: Sov.Math.Usp. {\bf 37}, 3 (1982)}

\lref\pw{Polyakov,A.M., Wiegmann,P.B.: Goldstone fields in two
dimensions with multi-valued action. Phys.Lett. {\bf 141}B, 223 (1984)}

\lref\bpz{Belavin,A.A., Polyakov,A.M., Zamolodchikov,A.B.: Infinite
conformal symmetry in two-dimensional quantum field
theory. Nucl.Phys. {\bf 241}B, 333 (1984)}

\lref\kz{Knizhnik,V.G., Zamolodchikov,A.B.: Current algebra and
Wess-Zumino model in two dimensions. Nucl.Phys. {\bf 247}B, 83 (1984)}

\lref\wg{Gepner,D. Witten,E.: String theory on group
manifolds. Nucl.Phys. {\bf 278}B, 493 (1986)}

\lref\zf{Fateev,V.A.,Zamolodchikov,A.B.: Operator algebra and
correlation functions in the two-dimensional Wess-Zumino $SU(2) \times
SU(2)$ chiral model. Sov.J.Nucl.Phys {\bf 43}, 657 (1986)}

\lref\bud{Budagov,A.S., Takhtadzhyan,L.A.: A nonlinear one-dimensional
model of classical field theory with internal degrees of
freedom. Sov.Phys.Dokl. {\bf 22(8)}, 428 (1977)}

\lref\fad{Faddev,L.D., Takhtadzhyan,L.A.: Hamiltonian methods in the
theory of solitons. Berlin,Germany: Springer (1987) 592p. (Springer
series in soviet mathematics).}

\lref\vafa{Cecotti,S., Vafa,C.: Topological antitopological
fusion. Nucl.Phys. {\bf 367}B, 359 (1991)}

\lref\fs{Faddev,L.D., Takhtadzhyan,L.A.: Essentially nonlinear
one-dimensional model of the classical field
theory. Theor.Math.Phys. {\bf 21}, 1046 (1975)}

\lref\zm{Shabat,A.B., Zakharov,V.E.: Func.Anal.Appl. {\bf 13}, 13
(1979)}

\lref\zz{Zamolodchikov,A.B., Zamolodchikov,Al.B.: Factorized S
matrices in two-dimensions as the exact solutions of certain
relativistic quantum field models. Ann.Phys. {\bf 120} 253 (1979)}

\lref\bh{Witten,E.:On string theory and black holes. Phys.Rev. {\bf
44}D, 314 (1991)}

\lref\znfat{Fateev,V.A.:Integrable deformations in $Z_n$ symmetrical
models of conformal field theory.  Int.J.Mod.Phys.{\bf 6}A, 2109 (1991)}

\lref\pohl{Pohlmeyer,K.:Integrable hamiltonian systems and
interactions through quadratic constraints. Comm.Math.Phys. {\bf 46},
207 (1976)}

\lref\za{Zamolodchikov,A.B.: Integrable field theory from conformal
field theory. Adv.Stud. Pure.Math. {\bf 19}, 641 (1989)}

\newsec{Introduction.}

Conformal field theory and integrable field theory in two dimensions
are two subjects which attracted a lot of attention in the last years.
The reason essentially lies in specific two-dimensional symmetries
which lead to exact solutions of the quantum field dynamics. In
conformal field theory solvability of a model is provided by a chiral
algebra,  which includes Virasoro algebra as subalgebra. Examples
are W-algebras \wz \lf , parafermions \parzf  \lw , Wess-Zumino
model \wzwb , etc.   When we go away from criticality all the symmetry is usually
lost. However, as it was pointed out in \zaza , for certain classes of 
perturbations of a conformal field theory the resulting
two-dimensional  field theory possesses infinite number of mutually
commutative conserved charges. The currents corresponding to these
charges can be  viewed as a deformation of some operators from
(enveloping of) chiral algebra. The conserved charges result in a
drastically  simplified scattering theory -
a general S matrix factorizes into product of two-particle S
matrices \kul  \kar . Factorized scattering preserves the number of particles and
the set of on-mass-shell momenta. This great simplification
makes it possible  to probe a  vicinity of a conformal fixed
point in detail  \zaal . Large variety of theories with factorized scattering  was
explicitly constructed  \musa \fre  \zhu  \musb  \she  \lecla . Many of them are  
reductions \leclb  \smir   at a special value of coupling constants  of quantized
classical field theories,  e.g. Toda field theories. The factorization
is  typical for the scattering of solitons of the  nonlinear classical field equations
integrable by the inverse scattering method. The spectra of conserved
charges  in such a theories are  essentially
the same as in the  reduced versions. Factorized scattering shows up
in semiclassical limit as absence of backward scattering of lumps.
Most of the considered models  contain one or few stable particles and
exhibit no resonances.

 In this paper we consider a model containing a bunch of unstable
particles, besides a stable ones. From 
the  point of view of conformal field theory the this model is a
perturbation of  $SU(2)_k$ Wess-Zumino-Witten
model by a certain multiplet of primary operators. The perturbation depends in
general on four arbitrary parameters. The layout of the paper follows. In
section 2 we introduce the necessary notations and recall some basic facts
about Wess-Zumino-Witten model. In section 3 we formulate the model,
obtain recurrence formula for classical conserved currents and
construct $N$-kink solution for the model describing scattering, merge
and decay of kinks and their bound states. Section 4 is devoted to the
construction of quantum conserved charges. Besides the series of
integrals of motion which survive semiclassical limit there is an 
additional set of conserved charges, not admitting semiclassical
limit.  A brief consideration of scattering amplitudes is
given in section 5

\newsec{Basics.}

In this section we recall basic facts about Wess-Zumino-Witten model
and  introduce necessary the notations. 

Consider two-dimensional Wess-Zumino-Witten (WZW) action \wzwa , \wzwb ,\wzwc
\eqn\swzw{{\cal S}_{wzw}\,=\,{-k \over {16 \pi}} \int d t d x\, {\rm Tr}\,\(\d^{\mu} g
g^{-1}\,\d_{\mu} g g^{-1}\)\,+\,k \Ga \[g\]\,}
where field $g(t,x)$ takes value in any semisimple compact group
$G$, and  $t, x$ - are coordinates of 2-dimensional Euclidean space
($t$ is euclidean time) and $k$ is some dimensionless coupling constant. 
Here we will deal mostly with $G=SU(2)$.  The ambiguous
term $\Ga\[g\]$  is given by the functional
\eqn\G{\Ga \[\tl g\] \,=\,{1\over {24 \pi}} \int
d t d x d \tau\,\ep_{\mu \nu \la} {\rm Tr}\, \(\d^{\mu} \tl g \tl g^{-1}\,\d^{\nu}
\tl g \tl g^{-1}\,\d^{\la} \tl g \tl g^{-1}\)\,}
where integration is over 3-dimensional half-space $(t,x,\tau)$, $\tau
\ge 0$. Boundary conditions $\tl g(t,x,\infty)=1$, $\tl
g(t,x,0)=g(t,x)$ define the functional $\Ga$ modulo $2 \pi N$ for some
integer $N$. With these boundary conditions functional integral is
well-defined if $k$ is a positive integer number. It shown in \wzwb
 \pw   the model \ \swzw\ is conformally invariant and therefore can
be  studied by methods of two-dimensional conformal field theory (CFT)
\bpz . In fact this model
possesses larger symmetry with respect to current algebra
$\wtl{su(2)_{k}} \times \wtl{su(2)_{k}}$ \pw , \kz . Namely, the action
\ \swzw\ is  invariant under transformations 
\eqn\gtr{g\,\rightarrow \,\Om(z)\, g(z,\bar z)\, \bar \Om(\bar z)\,}
where $\Om(z)$ and $\bar\Om(\bar z)$ are arbitrary $SU(2)$-valued
functions of light-cone variables
\eqn\lc{z\,=\,(t + i x)/2,\,\,\,\,\,\bar z\,=\,(t - i x)/2\,}

The field content, anomalous dimensions of the fields and
equations for correlation function were studied in \kz . Partition
functions and applications to string theory were discussed in \wg . 
The infinite dimensional symmetry \ \gtr\ of the WZW model is
generated by local currents
\eqn\cur{J\,=\,-\frac{k}{4} \d g g^{-1},\,\,\,\,\,\,\bar
J\,=\,-\frac{k}{4} g^{-1} \bar \d g\,}
which satisfy equations \wzwb 
\eqn\dcur{\bar \d J\,=\,\d \bar J\,=\,0\,}
The currents $J(z)$ and $\bar J(\bar z)$ obey the following operator
product expansion (OPE) \kz
\eqn\opecur{J^{a}(z_{1})\,J^{b}(z_{2})\,=\,\frac{k}{2}
\frac{q^{ab}}{(z_{1}-z_{2})^2}\,+\, \frac{f^{ab}_{c}}{(z_{1}-z_{2})}
J^{c}(z_{2})\,+\,{\rm regular\,\,\, terms}\,}
here tensors $q^{ab}$ and $f^{ab}_{c}$ are $su(2)$ invariant metric
and structure constants. They have components
\eqn\strc{q^{00}\,=\,\half q^{+-}\,=1,\,\,\,\,\,f^{+-}_{0}\,=\,2,\,\,\,\,\,
f^{0+}_{+}\,=\,-f^{0-}_{-}\,=\,1\,}
Let $\cal F$ be  space of mutually local fields of the theory. For any
$F(z,\bar z)\,\in \cal F$ we define an action of operators $J^{a}_{n}$,
($a=\pm,0$ and $n=0,\pm 1,\pm 2,..$) acting in this space
\eqn\curact{J_n^{a}F\,(z,\bar z)\,=\,\oint_{{\cal C}_{z}}\,\frac{d \zeta}{2 \pi i}
J^{a}(\zeta) (\zeta - z)^{n} F(z,\bar z)\,}
As it follows from \ \opecur\ operators $J^{a}_{n}$ satisfy commutation
relations of $\wtl {su(2)_k}$ Kac-Moody algebra
\eqn\kmcom{\[J^{a}_{n},J^{b}_{m}\]\,=\,f^{ab}_{c}\,J^{c}_{n+m}\,-\,\frac{km}{2}
q^{ab} \de_{n+m,0}\,}
Since  the theory has conformal invariance energy-momentum tensor is
traceless and has only two independent components
\eqn\tdef{\eqalign{&T(z)\,=\,\frac{1}{(k+2)} q_{ab}\, :J^{a}(z) J^{b}(z):\cr 
&\bar T(\bar z)\,=\,\frac{1}{(k+2)} q_{ab}\, :\bar J^{a}(\bar z) \bar
J^{b}(\bar z):\cr}\,}
Field $T(z)$ satisfy OPE
\eqn\opett{T(z_{1})\,T(z_{2})\,=\,\frac{c}{2}
\frac{1}{(z_{1}-z_{2})^4}\,+\,\frac{2
T(z_{2})}{(z_{1}-z_{2})^2}\,+\,\frac{\d
T(z_{2})}{(z_{1}-z_{2})}\,+\,{\rm regular\,\,\,terms}\,}
Again as in the case with Kac-Moody algebra we can define action of
generators $L_{n},\,\,n=0,\pm 1,\pm 2,..$ in the space $\cal F$
\eqn\tact{L_{n}F\,(z,\bar z)\,=\,\oint_{{\cal C}_{z}} \,\frac{d \zeta}{2 \pi i}
T(\zeta) (\zeta - z)^{n+1} F(z,\bar z)\,}
Operators $L_{n}$ form Virasoro algebra
\eqn\vircom{\[L_{n},L_{m}\]\,=\,(n-m) L_{n+m}\,+\,\frac{c}{12} (n^3-n)
\de_{n+m,0}\,}
with central charge 
\eqn\cc{c=3 k/(k+2)\,}
and commute with Kac-Moody generators $J^{a}_{m}$ as
\eqn\comlj{\[J^{a}_{m},L_{n}\]\,=\,m J^{a}_{m} \,} 
In terms of $J^{a}_{n}$ Virasoro generators $L_{n}$ can be expressed as
\eqn\lmd{L_{n}\,=\,\frac{q_{ab}}{(k+2)}\sum_{m=-\infty}^{+\infty}
:J^{a}_{m} J^{b}_{n-m}:\,}
where symbol $:(\cdots):$ means  normal ordering, i.e. operators $J^{a}_{n}$
with $n<0$ to be put to the left. Operators $\bar
J^{a}_{n}$ and $\bar L_n$ related to the fields $\bar J^{a}(\bar z)$,
$\bar T(\bar z)$ can be defined in a similar way .
There are invariant fields \zf , \kz
$\Phi^{(j \bar j)}_{m,\bar m},\,\,j,\bar j\bar j=0,1,2,..,k/2,\,\,
m,\bar  m=-j,-j+1,..,j-1,j$ in the space $\cal F$ which obey  equations
\eqn\invf{\eqalign{&J^{a}_{n} \Phi^{(j,\bar j)}_{m,\bar m}\,=\,0,\,\,\,\,\,\,n>0\cr
&J^{+}_{0}\Phi^{(j,\bar j)}_{m,\bar m}\,=\,
\[(j-m)(j+m+1)\]^{1/2}\,\Phi^{(j,\bar j)}_{m+1,\bar m}\cr
&J^{0}_{0}\Phi^{(j,\bar j)}_{m,\bar m}\,=\,
m\,\Phi^{(j,\bar j)}_{m,\bar m}\cr
&J^{-}_{0}\Phi^{(j,\bar j)}_{m,\bar m}\,=\,
\[(j+m)(j-m+1)\]^{1/2}\,\Phi^{(j,\bar j)}_{m-1,\bar m}\cr}\,}
and the ``bar''-counterpart of the above equations.
Invariant fields  $\Phi^{(j,\bar j)}_{m,\bar m}$ have dimensions
$\(\De_{j},\De_{\bar j}\)$,  
\eqn\dimf{\De_{j}\,=\,\frac{j(j+1)}{k+2}\,}
If $j=\bar j$ we will denote the corresponding field as $\Phi^{(j)
}_{m,\bar m}$. The space $\cal F$ can be represented as a sum
\eqn\fexp{{\cal F}\,=\,\oplus_{j,\bar j}\[\Phi^{(j,\bar j)}\]\,}
where $\[\Phi^{(j,\bar j)}\]$ is a space spanned by fields of the form
\eqn\vect{J^{a_1}_{n_1} J^{a_2}_{n_2}...J^{a_s}_{n_s}\bar J^{b_1}_{m_1} \bar
J^{b_2}_{m_2}...  \bar J^{b_l}_{m_l}\Phi^{(j,\bar j)}(z,\bar z)\,}
In semiclassical limit $k \rightarrow \infty$ fields $\Phi^{(j,j)}$
become an $SU(2)$ matrix $g^{(j)}(z,\bar z)$ in representation of spin $j$
while fields \ \vect\ become differential polinomial in matrix
elements of $g^{(j)}$.

\newsec{Perturbation of WZW model. Semiclassical analysis.}

Now we turn to perturbation of WZW model. As a perturbation we take
operator $\cal O$
\eqn\perto{{\cal O}\,=\,\frac{1}{2 \pi} \int d t d x\,\BL {\rm Tr}\(g A
g^{-1} B\)\,-\,{\rm Tr}\(A B\) \BR\,}
In this section we make rotation $t \rightarrow - i t$ to Minkowski space.
The perturbed action functional become
\eqn\per{{\cal A}\,=\,{\cal S}_{wzw}\,+\,{\cal O}\,}
here $A$ and $B$ are diagonal matrices
\eqn\defab{A\,=\,{\rm
Diag}\(a_{1},a_{2},a_{3}\),\,\,\,\,\,B\,=\,{\rm
Diag}\(b_{1},b_{2},b_{3}\) \,}
and we take $g$ in real spin $j=1$ representation. Equivalently we can
say that $g$ takes values in $SO(3)$ instead of $SU(2)$.
If we introduce coordinates $\psi, \vth, \vphi$ on $SO(3)$ group manifold 
\eqn\grpar{g\,=\,\(\matrix{\cos \psi &\sin \psi&0\cr -\sin
\psi&\cos \psi &0\cr 0&0&1\cr}\) \(\matrix{\cos \vphi&0&\sin \vphi\cr 0&1&0\cr
-\sin \vphi &0&\cos \vphi \cr}\) \(\matrix{1&0&0\cr
0&\cos \vth &\sin \vth \cr 0&-\sin \vth &\cos \vth \cr}\)\,}
the action functional $\cal A$ takes the  form
\eqn\aexp{{\cal A}\,=\,\frac{k}{4 \pi}\int d x ^{+} d x^{-}\BL \d_{+} \psi\, {\d}_{-}
\psi\,+\,\d_{+} \vphi\, {\d}_{-} \vphi\,+\, \d_{+} \vth\, {\d}_{-} \vth\,-\, 2 \sin \vphi \d_{-}
\vth\, \d_{+} \psi \,-\,\ka^{2}\,U(\psi,\vphi,\vth)\BR\,}
where the potential $U$ is of extremely complicated form
\eqn\v{\eqalign{U\,=&\,\al_{1} \b_{1} \sin^{2} \psi\,+\,\al_{2} \b_{2} \sin^{2}
\vth\,+\al_{3} \b_{3} \sin^{2} \vphi\,+\cr
& \al_{2} \b_{1} \sin^{2} \psi \sin^{2} \vth \,-\,\al_{3}
\b_{1} \sin^{2} \psi \sin^{2} \vphi\,-\,\al_{2} \b_{3}
\sin^{2} \vth \sin^{2} \vphi\,+\cr 
&\al_{2} \b_{1} \sin \psi \sin \vphi \sin \vth \(2 \cos \psi \cos
\vth\,+\,\sin \psi \sin \vphi \sin \vth \)\,\,,\cr}\,}
$x^{\pm}=(t \pm x)/2$, $\ka^{2}\,=\,4/k$ and coupling constants defined as
\eqn\param{\eqalign{&\al_{1}=a_{1}-a_{2},\,\,\,\al_{2}=a_{2}-a_{3},\,\,\,\al_{3}=a_{1}-a_{3},\cr
&\b_{1}=b_{1}-b_{2},\,\,\,\b_{2}=b_{2}-b_{3},\,\,\,\b_{3}=b_{1}-b_{3}\cr}\,}

The dynamical model described by the above
action was  first considered in \bud . 

For general $\al_{i}$ and $\b_{i}$ the perturbation completely breaks $\wtl{su(2)_{k}}
\times  \wtl{su(2)_{k}}$ symmetry of the model. Another feature of the
model is completely broken $P$-parity , i.e symmetry with respect to
reflection of space coordinate $x \rightarrow -x$. WZW can be made invariant
under change of sign of $x$ if we simultaneously perform
transformation $g \rightarrow g^{-1}$. In the perturbed model we do
not have this option because of the specific form of the operator $\cal
O$. 

\subsec{Lax pair and classical integrals of motion (IM).}

In \bud   a Lax pair representation for \ \aexp\ was found and
it was shown that  the model can be solved using inverse scattering
method \fad .
For our purposes we will use another Lax pair which is more simple and
suitable for analysis. 

Equations of motion for the perturbed model \ \per\ 
\eqn\eqmo{\eqalign{&\d_{+} J\,=\,\[g A g^{-1},B\]\cr
&J\,\equiv\,-\frac{k}{4}\d_{-} g g^{-1}\cr}\,}
are compatibility condition for an auxiliary liner problem
\eqn\lin{\eqalign{&{\cal L}_{-} \Psi\,\equiv\,\BL \d_{-} \, -\, \d_{-} g g^{-1}\,-\, i
\la^{-1}\, \ka B  \BR \Psi\,=\,0 \cr
&{\cal L}_{+} \Psi\,\equiv\,\BL \d_{+} \,-\, i \la\, \ka g A g^{-1}\BR \Psi\,=\,0 \cr}\,}
where $\la$ is a spectral parameter. It is remarkable enough that both
the equation
\ \eqmo\ and the operators ${\cal L}_{\pm}$ (but without the 
spectral parameter $\la$) appeared also in a context of $N=2$
supersymmetric models  \vafa . 
The Lax pair \ \lin\ is a guaranty for existence of infinite number of
conserved charges 
\eqn\cons{\eqalign{&{\cal Q}_{a}^{(s)}\,=\,\int \BL Q_{a}^{(s)} d
x^{-}\,+\,R_{a}^{(s)} d x^{+}
\BR\cr &\dd{}{t} {\cal Q}_{a}^{(s)}\,=\,0\cr}\,}
in the model  \ \per\ (at least on classical level). In ${\cal
Q}_{a}^{(s)}$  $s$ refers to the spin of the charge and $a$
distinguish between charges of the same spin. Using the Lax
pair we find the recursion formula allowing to construct the
densities  $Q_{a}^{(s)}$
\eqn\rch{Q_{a}^{(s)}\,=\,\(J K^{(s)}\)_{aa}\,}
where matrix $K^{(s)}$ defined recursively
\eqn\rech{\eqalign{&K^{(s)}_{aa}\,\equiv\,0\cr
&i\ka (b_{m}-b_{n}) K^{(s+1)}_{mn}\,=\,\d_{-}
K^{(s)}_{mn}\,+\,\sum_{s_{1}+s_{2}=s} K^{(s_{1})}_{mn}
Q^{(s_{2})}_{n}\,-\,\(J K^{(s)}\)_{mn}\,-\,J_{mn} \de_{s,0}\cr}\,}
with initial conditions
\eqn\inc{K^{(0)}\,=\,0,\,\,\,\,\,Q^{(0)}\,=\,0\,}
A few first densities are
\eqn\ints{\eqalign{Q_{1}^{(1)}\,=&\,{\rm Tr}\(J J\),\cr
Q_{2}^{(1)}\,=&\,{\rm Tr}\(B J J\),\cr
Q_{a}^{(3)}\,=&\,-\sum_{p,q,s\ne a} \frac{J_{ap} J_{pq} J_{qs}
J_{sa}}{(b_{p}-b_{a}) (b_{q}-b_{a}) (b_{s}-b_{a})}\,+\,
\sum_{p \ne a} \frac{J_{ap} J_{pa}}{(b_{p}-b_{a})} \sum_{p \ne a}
\frac{J_{ap} J_{pa}}{(b_{p}-b_{a})^2}\,+\cr\,   
&2 \sum_{p,q \ne a} \frac{J_{aq} J_{qp} \d_{-}
J_{pa}}{(b_{q}-b_{a})(b_{p}-b_{a})^2}\,+\,\sum_{p \ne a} \frac{\d_{-}
J_{ap}\d_{-} J_{pa}}{(b_{p}-b_{a})^3}\cr}\,}
Only two of $Q_{a}^{(3)}$ are linearly independent. Further analisys
shows that allowed set of spins is $s\,=\,\pm 1,\pm 3,\pm 5,\pm 7,\pm 9,...$ and 
for any value of $s$ there are two linearly independent conserved
charges. Densities with negative $s$ can be obtained from the positive
ones replacing $J \,\rightarrow\,\frac{k}{4} g^{-1} \d_{+} g$ and
$b_{m}\,\rightarrow\,a_{m}$.  In the next section we construct
quantum conserved charges corresponding to \ \ints\ .

\subsec{Kinks.}

Now we are going to discuss soliton content of the model. First of all we
specify boundary conditions. The natural ones are $g \rightarrow
g_{\pm \infty}$ as $x \rightarrow \pm \infty$, with $g_{\pm \infty}$ being some
constant matrices. They can be determined from the condition of
vanishing of  the right hand side of \ \eqmo\ as $x \rightarrow \pm
\infty$.  There are four allowed vacua 
\eqn\vac{\eqalign{&g_{0}\,=\,{\rm Diagonal} \(1,1,1\)\cr
&g_{1}\,=\,{\rm Diagonal} \(-1,-1,1\)\cr
&g_{2}\,=\,{\rm Diagonal} \(1,-1,-1\)\cr
&g_{3}\,=\,{\rm Diagonal} \(-1,1,-1\)\cr}\,}
All these vacua have same energy. Therefore we expect that there are
kinks interpolating between them. Let for a moment $\vphi=0$ and
$\vth=0$ in \ \aexp\ . Then we get familiar sine-Gordon (sG) action for the
field $\psi$. It is well known that sG theory has kinks \fs , and the
corresponding solutions interpolate in our model between $g_{0}$ and
$g_{1}$.  The same happens if we set another couple of fields to zero.
However,  there is significant difference between sG kinks and kinks in
the model we are considering. Because
$\pi_{1}\(SO(3)\)\,=\,Z_{2}$ kinks must have $Z_{2}$ charge.
For example lets take kink $g(t,x)$ interpolating between $g_{0}$ and $g_{1}$. At
given moment $t$, $g(x)$ is a path on $SO(3)$ manifold connecting
identity element with $g_{1}$. But $g_{1}$ is nothing but rotation by angle
$\pm \pi$ and kinks with different $Z_{2}$ charge correspond to different
choice of the sign. On the contrary, sG kinks have  $Z$ charge.

Probably the easiest way to construct kink solutions is to use
Riemann problem \zm  for the Lax pair \ \lin\ . In this approach
solitons are related to $\Psi$'s in \ \lin\ having simple
analytical properties in $\la$. Namely,
\eqn\psya{\Psi(x^{-},x^{+}|\la)\,=\,\Pi(x^{-},x^{+}|\la) \cdot \exp\(i
\la^{-1}\ka B x^{-}\,+\,i \la \ka A x^{+}\)\,}
where $\Pi$ is a meromorphic function of $\la$. If we normalize $\Pi$ as
$\Pi(x^{-},x^{+}|0)\,=\,1$, then an example of such a  solution for $\Pi$ relevant for
scattering of kinks and having $N$ poles in $\la$-plane  can be written as
\eqn\psyasol{\[\Pi(\la)\]_{ab}\,=\,\frac{{\rm det\, M}^{(ab)}(\la)}{{\rm det\, M}}\,}
where matrices ${\rm M}^{(ab)}(\la)$ and ${\rm M}$ are $(N+1) \times (N+1)$ and
$N \times N$ correspondingly
\eqn\mdef{{\rm M}_{ij}\,=\,\frac{\ga_i\,\(e_i,e_j\)}{\ga_i+\ga_j},\,}
and
\eqn\mabdef{{\rm M}^{(ab)}(\la)\,=\,\(\matrix{
\de_{ab}&\frac{\la\,e_1^{(b)}}{\la+i\ga_1}
&\frac{\la\,e_2^{(b)}}{\la+i\ga_2}&\cdots &\frac{\la\,e_n^{(b)}}{\la+i\ga_n}\cr
e_1^{(a)}&{}&{}&{}\cr e_2^{(a)}&{}&{\rm M}&{}\cr \vdots
&{}&{}&{}\cr e_n^{(a)}&{}&{}&{}\cr}\)\,}
here $\ga_{i}$ are $N$ positive numbers (they are related to
rapidities $\th$ of kinks) and  $(e_{i},e_{j})$ is a scalar product of
vectors
\eqn\vecdef{e_{n}\,=\,\exp\(\,\ga_{n} \ka A x^{+}\,-\,\ga_{n}^{-1} \ka B x^{-}\) \cdot
c_{n},\,}
here $c_{n}$ are some real 3-dimensional vectors independent of
$z$ and $\bar z$. 

The corresponding solution for $g$ is
\eqn\gsol{g(x^{-},x^{+})\,=\,g_{a} \cdot \Pi(x^{-},x^{+}|\infty) \cdot g_{b}\,}
and the current $J$ is determined by expansion of $\Pi$ near $\la=0$
\eqn\jsol{ \[\Pi(x^{-},x^{+}|\la)\]_{mn}\,=\,\de_{mn}\,+\,i \la
\frac{\[J(x^{-},x^{+})\]_{mn}}{(b_{m}-b_{n})}\,+\,O(\la^2)\,}

The simplest solutions are kinks $K_{0a}^{(\ep)}(\th)$ with topological
charge $\ep$ interpolating
between $g_{0}$ and $g_{a}$. They correspond to the case $N=1$ and one of
components of the vector $c$ is zero in \ \gsol\ . For example, for $K_{01}^{(\ep_{1})}(\th)$
\eqn\kink{g\,=\,\(\matrix{-\frac{\sinh \phi_{1}}{\cosh \phi_{1}} &
\frac{(-1)^{\ep_1}}{\cosh \phi_{1}} & 0\cr -\frac{(-1)^{\ep_1}}{\cosh
\phi_{1}} & -\frac{\sinh \phi_{1}}{\cosh \phi_{1}} & 0\cr 0 & 0 & 1\cr}\)\,}
here 
\eqn\faza{\phi_{1}\,=\,\frac{\ka}{2} \(\al_{1} \ga\,+\,\b_{1}
\ga^{-1}\) x\,+\,\frac{\ka}{2} \(\al_{1} \ga\,-\,\b_{1} \ga^{-1}\)
t\,+\,\de_{1}\,}
$\ep_{1}=0,1$ is $Z_{2}$ topological charge of the kink and $\de_{1}$
is an arbitrary real constant.
We see that only field $\psi$ is exited. There are similar
solutions for kinks $K_{02}^{(\ep_{2})}(\th)$ and  $K_{03}^{(\ep_{3})}(\th)$
when only fields $\vth$ or $\vphi$ are excited. Using the energy-momentum
tensor we find that kinks $K_{0a}^{(\ep)}(\th)$ and ones obtained from
them by multiplication by $g_{n}$'s ( see \ \vac\ ) have two-dimensional
momenta 
\eqn\mom{P^{\mu}_{(n)}\,=\,\(M_{n} \cosh \th_{n}, M_{n} \sinh
\th_{n}\)\,}
where masses of kinks
\eqn\mass{M_{n}\,=\,\frac{1}{\pi} \( k\, \al_{n}\b_{n}\)^{1/2}\,}
and rapidities $\th_{n}$ are related to $\ga$ 
\eqn\rap{\th_{n}\,=\,-\,\ln \ga \,+\,\half \ln \(\b_{n}/\al_{n}\)\,}
now we can rewrite phase \ \faza\ as
\eqn\phya{\phi_{1}\,=\,\frac{2 \pi M_{1} }{k} \(x \cosh \th_{1}\,-\,t
\sinh \th_{1}\)\,+\,\de_{1}\,}
In the following we will assume that
\eqn\relpar{a_{1}\,>\,a_{2}\,>\,a_{3},\,\,\,\,\,\,\,\,\,\,
b_{1}\,>\,b_{2}\,>\,b_{3}\,}

In general we have $3 N$ free parameters in \ \gsol\ . Restricting
components of $c_{i}$ by condition $c_{i}^{(1)}\cdot c_{i}^{(3)}
=0,\,\,i=1,2,..,N$  we obtain $2 N$
parametric solution describing scattering of kinks. We give an example of
scattering of two different kinks
$K_{01}^{(\ep_1)}(\th_{1})\,K_{13}^{(\ep_2)}(\th_{2})\,
\rightarrow\,K_{02}^{(\ep_2)}(\th_{2})\,K_{23}^{(\ep_1)}(\th_{1}) $
($N=1,\,\,\,c_{1}^{(3)}=0,\,\, c_{2}^{(1)}=0$). From \ \grpar\ and \
\gsol\ we find, for example
\eqn\tgps{\eqalign{&\tan \psi\,=\,-(-1)^{\ep_{1}}\frac{2 e^{\phi_{1}} (1+\De e^{2
\phi_{2}})}{1-e^{2 \phi_{1}}+\De^{2} e^{2 \phi_{2}}-e^{2 (\phi_{1}+
\phi_{2})}}\cr
&\sin \vphi\,=\,(-1)^{\ep_{1}+\ep_{2}}\frac{2 (\De-1) e^{\phi_{1}+\phi_{2}}}
{1+e^{2 \phi_{1}}+\De^{2} e^{2 \phi_{2}}+e^{2 \phi_{1}} e^{2 \phi_{2}}}\cr
&\tan \vth\,=\,-(-1)^{\ep_{2}}\frac{2 (e^{2 \phi_{1}}+\De ) e^{\phi_{2}}}
{1+e^{2 \phi_{1}}-\De^{2} e^{2 \phi_{2}}-e^{2
(\phi_{1}+\phi_{2})}}\cr}\,}
here 
\eqn\del{\De\,=\,\tanh \(\frac{\th_{1}-\th_{2}}{2}\,-\,\frac{1}{4} \ln
\(\frac{\al_{1} \b_{2}}{\al_{2} \b_{1}}\)\),\,} 
\eqn\phyaa{\phi_{n}\,=\,\frac{2 \pi M_{n}}{k} \(x \cosh \th_{n}\,-\,t
\sinh \th_{n}\)\,+\,\de_{n}\,}
In full form the corresponding solution is given in Appendix A.

What are the rest of parameters responsible for? In general solution \
\gsol\ describes scattering, merge and decay of solitons. The simplest
example is decay process $K_{03}^{(\ep_{1} + \ep_{2})}(\th_{3})
\rightarrow K_{02}^{(\ep_{2})}(\th_{2})\, K_{23}^{(\ep_{1})}(\th_{1})$
($N=1$, all components of $c_{1}$ are nonzero) 
\eqn\decex{g\,=\,\(\matrix{\frac{1+e^{2 \phi_{2}}-e^{2 (\phi_{1} +\phi_{2})}}{1+e^{2
\phi_{2}}+e^{2 (\phi_{1}+\phi_{2})}} & 
\frac{-2 (-1)^{\ep_{1}} e^{\phi_{1} +2 \phi_{2}}}{1+e^{2 \phi_{2}}+e^{2 (\phi_{1}+ \phi_{2})}}&
\frac{-2 (-1)^{\ep_{1}+\ep_{2}} e^{\phi_{1} + \phi_{2}}}{1+e^{2\phi_{2}}+D_{m,n}
e^{2 (\phi_{1}+\phi_{2})}}\cr
\frac{-2 (-1)^{\ep_{1}} e^{\phi_{1} +2 \phi_{2}}}{1+e^{2 \phi_{2}}+e^{2 (\phi_{1}+\phi_{2})}}&
\frac{1-e^{2 \phi_{2}}+e^{2 (\phi_{1}+\phi_{2})}}{1+e^{2 \phi_{2}}+e^{2 (\phi_{1}+\phi_{2})}}& 
\frac{-2 (-1)^{\ep_{2}} e^{\phi_{2}}}{1+e^{2 \phi_{2}}+e^{2 (\phi_{1}+\phi_{2})}}\cr
\frac{2 (-1)^{\ep_{1}+\ep_{2}} e^{\phi_{1} + \phi_{2}}}{1+e^{2 \phi_{2}}+e^{2 (\phi_{1}+\phi_{2})}}&
\frac{2 (-1)^{\ep_{2}} e^{\phi_{2}}}{1+e^{2 \phi_{2}}+e^{2 (\phi_{1}+\phi_{2})}}&
\frac{1-e^{2 \phi_{2}}-e^{2 (\phi_{1}+\phi_{2})}}{1+e^{2
\phi_{2}}+e^{2 (\phi_{1}+\phi_{2})}}\cr}\),\,}
with phases $\phi_{n}$
\eqn\pha{\phi_{n}\,=\,\frac{\ka}{2} \(\al_{n} \ga\,+\,\b_{n}
\ga^{-1}\) x\,+\,\frac{\ka}{2} \(\al_{n} \ga\,-\,\b_{n} \ga^{-1}\)
t\,+\,\de_{n}\,}
Indeed, suppose that $\al_{3} \ga\,-\,\b_{3} \ga^{-1}\,=\,0$, $\al_{2}
\ga\,-\,\b_{2} \ga^{-1}\,>\,0$,   and consider worldline
$\phi_{1}+\phi_{2}\,=$ const.  Then if $t \rightarrow
-\infty$ 
\eqn\limd{g\,\longrightarrow\,\(\matrix{\frac{1-e^{2 (\phi_{1}
+\phi_{2})}}{1+e^{2 (\phi_{1}+\phi_{2})}} & 0 &
\frac{-2 (-1)^{\ep_{1}+\ep_{2}} e^{\phi_{1} + \phi_{2}}}{1+e^{2 (\phi_{1}+\phi_{2})}}\cr
0&1&0\cr
\frac{2 (-1)^{\ep_{1}+\ep_{2}} e^{\phi_{1} + \phi_{2}}}{1+e^{2 (\phi_{1}+\phi_{2})}}& 0 &
\frac{1-e^{2 (\phi_{1}+\phi_{2})}}{1+e^{2
(\phi_{1}+\phi_{2})}}\cr}\)\,}
i.e. at $t=-\infty$ we have resting kink $K_{03}^{(\ep_{1} + \ep_{2})}$. Now consider limit $t
\rightarrow +\infty$. We first keep $\phi_{1}\,=\,$ const. In this
limit we get
\eqn\limdd{g\,\longrightarrow\,\(\matrix{\frac{1-e^{2 \phi_{1}}}{1+e^{2 \phi_{1}}} & 
\frac{-2 (-1)^{\ep_{1}} e^{\phi_{1}}}{1+e^{2 \phi_{1}}} & 0 \cr
\frac{-2 (-1)^{\ep_{1}} e^{\phi_{1}}}{1+e^{2 \phi_{1}}} &
\frac{-1+e^{2 \phi_{1}}}{1+e^{2 \phi_{1}}} & 0 \cr
0 & 0 & -1 \cr}\)\,}
that is $K_{23}^{(\ep_{1})}$ kink propagating in the positive $x$ direction.
Taking limit with  $\phi_{2}\,=\,$ const we obtain $K_{02}^{(\ep_{2})}$ kink
propagating in the negative $x$ direction.
\eqn\limmd{g\,\longrightarrow\,\(\matrix{
1 & 0 & 0\cr
0 & \frac{1-e^{2 \phi_{2}}}{1+e^{2 \phi_{2}}} & 
\frac{-2 (-1)^{\ep_{2}} e^{\phi_{2}}}{1+e^{2 \phi_{2}}}\cr
0 & \frac{2 (-1)^{\ep_{2}} e^{\phi_{2}}}{1+e^{2 \phi_{2}}}&
\frac{1-e^{2 \phi_{2}}}{1+e^{2 \phi_{2}}}\cr}\),\,}
That instability of the heaviest kink $K_{03}$ can be also predicted from
the mass spectrum of solitons \ \mass\ 
\eqn\maenq{M_{3}\,\ge\,M_{1}\,+\,M_{2}\,}
where equality achieved when $A=B$ in \ \per\ . The solution \ \decex\
can also be obtained from \ \tgps\ in a limit $\ga_{1} \rightarrow \ga_{2}$.

Now we have a very peculiar situation: the model has
infinite number of local conserved charges \ \cons\ and at the same time
there are unstable particles. If all particles were stable the
standard argument \zz  relate the presence of infinite number of
local conserved charges to factorization  of a scattering matrix into
two-particle $S$ matrices. In quantum theory unstable particles are
usually  excluded from ``in'' and ``out'' states, and the only trace
of them are resonance poles in $S$-matrix of the
theory. From this point of view it is crucial to find quantum
counterpart of charges \ \cons\ to discuss quantum integrability of \
\per\ , $S$-matrix, etc.  We address this problem in the next
section. 

As in sG model  there are bound states of kinks.  Solutions describing
bound states, scattering processes of bound states and kinks can be
obtained from \ \gsol\ by replacement of  some poles $\ga_{n}$ by pair of
poles $\ga_{n} \pm i \varepsilon_{n}$.

\subsec{$\wtl {so(2)}$ reduction of the model.}

One can note that when one of $\al_{n}$ or $\b_{n}$
vanish, the model regains right or left $\wtl {so(2)}$ symmetry.
If, for example, $\al_{3}$ and $\b_{3}$ vanish simultaneously $\wtl
{so(2)} \times \wtl {so(2)}$ symmetry is regained in the model. It
means that there is a propagating massless field. It is possible to
factor out this mode as in \bh  , \pohl   and get another model described by
the functional \footnote{$^*$}{This model is closely related to
$Z_{n}$ parafermionic models perturbed by second thermal operator
$\varepsilon _{2}$  \znfat  .}
\eqn\bh{{\cal S}\,=\,\int d^2 z\BL\,\frac{\d^{\mu}u\,\d_{\mu}{\bar
u}}{1-u \bar u}\,+\,2 \la \((u \bar u)^2-u \bar u\)\,\BR,\,}
which after change of variables 
\eqn\chv{u\,=\,\exp \(\frac{i \al}{2}\) \sin
\(\frac{\th}{2}\),\,\,\,\,\,\,\bar u\,=\,\exp \(-\frac{i \al}{2}\) \sin
\(\frac{\th}{2}\),\,}
takes the  form
\eqn\bhm{{\cal S}\,=\,\frac{1}{4} \int d^2 z \BL\,\d^{\mu} \th \d_{\mu}
\th\,+\,\tan^{2} \(\th/2\)\,\d^{\mu} \al \d_{\mu} \al\,+\,\la
(\cos 2 \th \,-\,1)\,\BR\,} 
This model is integrable and possesses conserved charges of spin
$s\,=\,\pm 1,\pm 3,\pm 5,...$. To describe the densities of these
charges we introduce new fields $\om$ and $\ga$ \pohl  which
satisfy  equations
\eqn\eqom{\eqalign{&\d \om\,=\,\d \al \frac{\cos \th}{2 \cos^{2}
\frac{\th}{2}},\,\,\,\,\,\bar \d \om\,=\,\bar \d \al \frac{1}{2 \cos^{2}
\frac{\th}{2}} \cr
& \d \ga\,=\,\d \al \frac{1}{2 \cos^{2}
\frac{\th}{2}},\,\,\,\,\,\bar \d \ga\,=\,\bar \d \al \frac{\cos \th}{2 \cos^{2}
\frac{\th}{2}}\cr}\,} 

The above equations are compatible if equations of motion of \ \bhm\
are being imposed. Using $\om$ and $\ga$ we construct currents
\eqn\curlr{\eqalign{&j^{\pm}\,=\,\( \d \th\,\pm\,i\, \d \al \tan \frac{\th}{2}\)
e^{ \pm i \om}\cr
&\bar j^{\pm}\,=\,\( \bar \d \th\,\pm\,i\, \bar \d \al \tan \frac{\th}{2}\)e^{ \pm i
\ga}\cr}\,}
For $\la=0$ equations of motion for \ \bhm\ are equivalent to
\eqn\concurbh{\bar \d j^{\pm}\,=\,\d \bar j^{\pm}\,=\,0\,}
Local integrals of motion (IM)  for \ \bhm\ turn out to be differential polynomials   in
$j^{\pm}$ and $\bar j^{\pm}$.   The first ones, $s=\pm 1, \pm 3$  are

\eqn\intbh{\eqalign{&Q^{(1)}\,=\,j^{+} j^{-}\cr
&\bar Q^{(1)}\,=\,\bar j^{+} \bar j^{-}\cr
&Q^{(3)}\,=\,\(j^{+} j^{-}\)^{2}\,+\,2\, \d j^{+}\, \d j^{-}\cr
&\bar Q^{(3)}\,=\,\(\bar j^{+} \bar j^{-}\)^{2}\,+\,2\, \bar \d \bar j^{+}\,
\bar \d \bar j^{-}\cr}\,}

\newsec{Quantum Integrals of Motion.}

In this section we will construct IMs proving
quantum integrability of the model discussed in section 3.

The central object of quantum theory are correlation functions  
of some operators ${\cal O}_{n}$ 
\eqn\corf{<{\cal O}_{1}(z_{1},\bar z_{1})\cdots {\cal
O}_{n}(z_{n},\bar z_{n})>_{\cal A} \, =\,\int {\cal
D} \phi\, {\cal O}_{1}(z_{1},\bar z_{1}) 
\cdots {\cal O}_{n}(z_{n},\bar z_{n})\, \exp\(-{{\cal A}\[\phi\]}\)\,}
where fields $\phi$ are some ``basic'' fields, 
operators ${\cal O}_{n}$ are some local 
functionals of $\phi$'s and their derivatives, and ${\cal A}\[\phi\]$
is an action functional. In this section we are back to euclidean two
dimensional space. In our case ${\cal A}$ is sum of WZW action 
\ \swzw \ $ {\cal S}_{wzw}$, which is conformally invariant and
represent  fixed point of renormalization group, and perturbation \
\perto\ . We will assume that the structure of space of fields in the
perturbed model is the same as in WZW model \ \fexp\ . To identify
operator $\cal O$ as an operator in WZW model we use \ \v\ to obtain
\eqn\oex{\eqalign{{\cal O}\,=\,-&\frac{1}{2 \pi} \int d^2
z\,\,\eta_{m}  \bar \eta_{\bar m}
\Phi^{(2)}_{m,\bar m}\(z, \bar z\) \, -\frac{1}{2 \pi} \int d^2
z\,\,\eta \cdot 1\,\cr 
&\equiv \,{\cal O}^{(2)}\,+\,{\cal O}^{(0)}\cr}\,}
where vectors $\eta^{m}$, $\bar \eta_{\bar m}$ and constant $\eta$ are
given by
\eqn\coex{\eqalign{&\eta_{m}\,=\,\(\frac{\b_{3}}{2},0,\frac{\b_{1}-\b_{2}}{\sqrt
6},0,\frac{\b_{3}}{2}\),\,\,\,\,\,\,\bar \eta_{m}\,=\,\(\frac{\al_{3}}{2},0,
\frac{\al_{1}-\al_{2}}{\sqrt 6},0,\frac{\al_{3}}{2}\),\cr 
&\eta\,=\,\frac{1}{3}{\rm Tr}A \cdot{\rm Tr}B\,-\,{\rm Tr}\( A B\)
\cr}\,}
For $k \ge 6$ operators $\Phi^{(2)}_{m,\bar m}$ are relevant and we
will show soon that the action is well defined for $k>10$ (see
discussion of `` resonance'' condition below) .
After expansion in ${\cal O}^{(2)}$ to the first order we get 
\eqn\pertexp{\eqalign{<{\cal O}_{1}(z_{1},\bar z_{1})\cdots {\cal
O}_{n}(z_{n},\bar z_{n})>_{\cal A}\,=\,&<{\cal O}_{1}(z_{1},\bar
z_{1})\cdots {\cal O}_{n}(z_{n},\bar z_{n})>_{wzw} e^{\frac{\eta}{2
\pi} V}\,-\cr
&<{\cal O}^{(2)}\,{\cal O}_{1}(z_{1},\bar z_{1})\cdots {\cal
O}_{n}(z_{n},\bar z_{n})>_{wzw}\,e^{\frac{\eta}{2 \pi} V}\, +\cdots\cr}\,} 
Here $<\cdots>_{wzw}$ means that average is taken with respect to WZW
action and $V$ is a volume of the base space. Note that the
perturbation theory around conformal fixed point is IR divergent and
to make it finite we fix $V$ finite but large. 
Here we  concentrate on equations of motion which are
relations between local operators and their derivatives. From
this point of view IR divergences are irrelevant.

Suppose we want to calculate $\bar \d$ derivative  of some holomorphic
(at fixed point) operator, say $Q(z)$, which is local with respect to $\Phi^{(2)}_{m,
\bar m}(z,\bar z)$. After taking derivative of \ \pertexp\ and
neglecting   usual contact terms (coming from the first term in \
\pertexp\ )  we get \za
\eqn\baseq{\bar \d Q(z,\bar z)\,=\,\eta_{m} \bar
\eta_{\bar m}\,\oint_{{\cal C}_{z}}  d \zeta\,\Phi^{(2)}_{m,\bar
m}\(\zeta, \bar z\)\, Q(z)\,}
If the residue of OPE 
\eqn\resope{{\rm Res}\BL \eta_{m} \bar \eta_{\bar m} \Phi^{(2)}_{m,\bar m}\(z, \bar z\) 
Q(w)\BR_{z=w}\,=\, \d R(w,\bar w)\,}
for some operator $R$, then we have constructed quantum conserved current 
\eqn\qint{\bar \d Q(z,\bar z)\,=\,\d R(w,\bar w)\,}
In general $\bar \d$ derivative of a local operator $Q$ in perturbed model
schematically take a form
\eqn\gendb{\bar \d Q\,=\,(\eta \bar \eta) R_{1}\,+\,(\eta \bar \eta)^2
R_{2}\,+\cdot \,}
where $R_{n}$ are some local fields. Because the structure of the
space of fields  in the perturbed WZW model was assumed
to be the same as in unperturbed model, we conclude that the series in
the right hand side of \ \gendb\ terminates at some term. Moreover,
all fields $R_{n}$ must have anomalous dimensions coinciding with the
WZW ones. If the anomalous dimension of some field $R_{n}$ matches the
one in WZW it must be added to the right hand side of \
\baseq\ . The ``resonance'' condition can be derived from \
\gendb\  by matching dimensions of operators on both sides of the
equation
\eqn\rescond{1\,-\,n\, (1-\De_{2})\,=\,\De_{j}\,}
for some $n>1$ and $j$. The solutions to the above equation
corresponds to additional term of order $\(\eta \bar \eta\)^{n}$ in \
\gendb\ . The only ``resonance'' cases are 
\eqn\ressol{\eqalign{&k\,=\,6\,\,\,\,\,\,\,\,\,j\,=\,0,\,\,\,n\,=\,4\,\,\,\,{\rm
and}\,\,\,\,j\,=\,1,\,\,\,n\,=\,3\cr
&k\,=\,7,\,\,\,\,\,\,\,\,\,j\,=\,0,\,\,\,\,n\,=\,3\cr
&k\,=\,8,\,\,\,\,\,\,\,\,\,j\,=\,1,\,\,\,\,n\,=\,2\cr
&k\,=\,10,\,\,\,\,\,\,\,\,\,j\,=\,0,\,\,\,\,n\,=\,2\cr}\,}
For that values of $k$ one need to consider next to the leading order
of the conformal perturbation expansion \ \pertexp\ to derive
equations of motion for local operators. Therefore we will restrict our attention to
the case $k>10$ when the action is well defined and will work within
the first order of the conformal perturbation theory.

\subsec{Coserved currents in $U\(\wtl {su(2)}_{k}\)$.}

Here we will find conserved currents which are composites of Kac-Moody
currents $J(z)$, \ \cur\ . 

Following \za  we define define new operators $D_{n,m},\,n=0,\pm 1,\pm
2,..,m=-2,...,2$ by its action in current module spanned by fields
$J^{a_1}_{n_1} J^{a_2}_{n_2}...J^{a_s}_{n_s} \cdot 1$
\eqn\deD{D_{m,n} Q\,(z,\bar z)\,=\,\oint_{{\cal
C}_{z}}  d \zeta\,\Phi^{(2)}_{m} \(\zeta, \bar z\)
\(\zeta\,-\,z\)^{n} Q(z)\,}
where notation $\Phi^{(2)}_{m}\(z, \bar z\)$ for the operator
$\eta_{\bar m} \Phi^{(2)}_{m,\bar m} \(z, \bar z\)$ was introduced.
Using \ \invf\ one can easily prove the commutations relations
\eqn\comd{\eqalign{&\[J^{+}_{p},D_{m,n}\]\,=\,\[(j-m)(j+m+1)\]^{1/2}
D_{m+1,p+n}\cr
&\[J^{0}_{p},D_{m,n}\]\,=\,m D_{m,p+n}\cr
&\[J^{-}_{p},D_{m,n}\]\,=\,\[(j+m)(j-m+1)\]^{1/2} D_{m-1,p+n}\cr}\,}
These operators satisfy simple relations
\eqn\prop{\eqalign{&\eta_{m} D_{m,0} Q\,(z)\,=\,-\bar \d Q(z,\bar z),\cr
&D_{m,-n-1} \cdot 1\,=\,\frac{1}{n!} L_{-1}^{n} \Phi^{(2)}_{m}(z,\bar z)\cr}\,} 
Now it is straightforward to find $\bar \d$ derivatives of any field from the
current module - apply $\eta_{m} D_{m,0}$ and then move all $D$'s all
the way to the right where they transforms into derivatives of
$\Phi^{(2)}_{m,\bar m}$. Sugavara construction \ \lmd\ allow to express
the result entirely in terms of Kac-Moody generators
$J^{a}_{n}$. Solving after that \ \qint\ we find quantum integrals of
motion  for spin $s=1,3$. For example
\eqn\qints{\eqalign{Q^{(1)}_{1}\,=\,\BL & J^{-}_{-1}
J^{+}_{-1}\,+\,J^{0}_{-1} J^{0}_{-1} \BR \cdot 1 \cr 
Q^{(1)}_{2}\,=\,\BL & \eta_{-2}\, J^{-}_{-1} J^{-}_{-1}\,-\,{\sqrt 6}
\eta_{0}\, J^{-}_{-1} J^{+}_{-1}\,+\, 
\eta_{2}\, J^{+}_{-1} J^{+}_{-1}\,\BR \cdot 1 \cr}\,}
The expressions for first two nontrivial quantum integrals of motion of spin
$s=3$ are given in Appendix B.

What we found is in perfect agreement with the classical result \
\ints\ . As before there are 2 conserved charges of spin $s=1$ and 2
charges of spin $s=3$. Charges with $s=-1$ and $s=-3$ can be obtained from
\ \qints\ and  replacing $J^{a}_{n} \rightarrow -
\bar J^{a}_{n}$ and $\eta_{m} \rightarrow \bar \eta_{m}$. It is
reasonable to make 

{\bf Conjecture.}  For general $k$ the model \ \per\ possesses infinite
number of mutually commutative local integrals of motion of spins
$s=0,\pm 1,\pm 2,...$  and for any $s$ there are two independent IM.

Note that the derivation does not feel if $k$ is integer or not.

\subsec{Additional set of  conserved charges for integer $k$.}

In this section for even $k$ we derive additional set of integrals of
motion.

The spectrum of anomalous dimensions of WZW theory \ \dimf\ contains
$\De_{\frac{k}{2}}\,=\,k/4$. It is
possible to show (see Appendix C) that the  symmetry algebra of WZW model can
be extended by new currents 
\eqn\newcur{\Psi_{m}(z)\,\equiv\,\Phi^{(\frac{k}{2},0)}_{m,0}(z),\,\,\,\,\,\,\,
{\rm and}\,\,\,\,\,\,\, \bar \Psi_{\bar m}(\bar
z)\,\equiv\,\Phi^{(0,\frac{k}{2})}_{0,\bar m}(\bar z)\,}
For $k=4n$, for example, these fields contribute to the partition
function of $SO(3)_{4n}$ WZW model, \wg . We are going to show that certain
linear combinations of $\Psi_{m},\,\bar \Psi_{\bar m}$ give rise to
new set of IM. 

To construct new IM's we need to consider OPE, \ \resope\ ,
\eqn\opeps{\eqalign{&\eta_{m} \bar \eta_{\bar m} \Phi^{(2,2)}_{m,\bar
m}(z_{1},\bar z_{1})\,\Psi_{n}(z_{2})\,=\cr
&C(2,\frac{k}{2},\frac{k}{2}-2)\,\eta_{m} \bar \eta_{\bar m}\(
\frac{\b_{m,n}^{l} \Phi^{(\frac{k}{2}-2,2)}_{l,\bar
m}(z_{2},\bar z_{2})}{(z_{1}-z_{2})^2}\,+\,\frac{\b_{m,n}^{a,l} J^{a}_{-1}
\Phi^{(\frac{k}{2}-2,2)}_{l,\bar m}\,(z_{2},\bar z_{2})}{(z_{1}-z_{2})}\,+\,\cdots\)\cr}\,}
Here $C(2,\frac{k}{2},\frac{k}{2}-2)$ is a structure constant of OPE (we will not
need its exact value), $\b_{m,n}^{l}$ and $\b_{m,n}^{a,l}$ are
some coefficients that can be determined comparing transformation
properties of the left and right hand sides of \ \opeps\ under action
of $\wtl {su(2)_{k}}$. To find $\b_{m,n}^{l}$ and
$\b_{m,n}^{a,l}$ we introduce following \zf  new variables $x,
\bar x$ and generating functions 
\eqn\gefu{\eqalign{&J(x,z)\,=\,J^{-}(z)\,+\,2 x\,J^{0}(z)\,-\,x^2\,J^{+}(z)\,\,\cr
&\Phi^{(j,\bar j)}(x,\bar x|z, \bar z)\,=\,\sum_{m=-j,\bar m=-\bar j}^{j,\bar
j}\,\[C^{m+j}_{2 j} C^{\bar m+\bar j}_{2 \bar j}\]^{1/2}\, x^{m+j}
\bar x^{\bar m+\bar j}\,\Phi^{(j,\bar j)}_{m,\bar m}(z,\bar z)\cr}\,}
where $C^{m+j}_{2 j} = (2 j)!/(j+m)!(j-m)!$ are binomial
coefficients. The singular part of the necessary OPE's \ \opecur\ take a form
\eqn\jj{\eqalign{&J(x_{1},z_{1})\,J(x_{2},z_{2})=-k
\frac{(x_{1}-x_{2})^2}{(z_{1}-z_{2})^2}\,- 2
\frac{x_{1}-x_{2}}{z_{1}-z_{2}}\,J(x_{2},z_{2})\,-
\frac{(x_{1}-x_{2})^2}{z_{1}-z_{2}}\,\frac{\d J(x_{2},z_{2})}{\d x_{2}}\cr
&J(x_{1},z_{1})\,\Phi^{(j,\bar j)}(x_2,\bar x_2|z_2, \bar z_2 )=-\BL 2 j
\frac{x_{1}-x_{2}}{z_{1}-z_{2}}\,+\,\frac{(x_{1}-x_{2})^2}{z_{1}-z_{2}}\,\frac{\d}{\d
x_{2}} \BR \,\Phi^{(j,\bar j)}(x_2,\bar x_2|z_2, \bar z_2)\cr}\,}
As in \ \curact\ we define action of modes $J^{a}_{n}(x)$
\eqn\accurn{J_{n}^{a}(x) F\,(x,\bar x| z,\bar z)\,=\,\frac{1}{\varepsilon_{a}
(2-a^2)} \oint_{{\cal C}_{z}} \,\frac{d \zeta}{2 \pi i}(\zeta -
z)^{n}\oint_{{\cal C}_{x}}\,\frac{d
y}{2 \pi i}(y-x)^{-2-a} J^{a}(y,\zeta) F(x,\bar x| z,\bar z)\,}
where $\varepsilon_{+}=\varepsilon_{0}=-\varepsilon_{-}=1$. OPE of
primary fields can be  written in a form
\eqn\opeff{\eqalign{&\Phi^{(j_{1},\bar j_{1})}(x_{1},\bar
x_{1}|z_{1}, \bar z_{1})\,\Phi^{(j_{2},\bar j_{2})}(x_{2},\bar
x_{2}|z_{2}, \bar z_{2} )\,=\, \sum_{j,\bar j} C \(\matrix{j_1 & j_2 & j \cr \bar j_1
&\bar j_2 &\bar j\cr}\) \cdot \cr
&(z_{12})^{\De_{j}-\De_{j_{1}}-\De_{j_{2}}}
(\bar z_{12})^{\De_{\bar j}-\De_{\bar j_{1}}-\De_{\bar j_{2}}}
(x_{12})^{{j}-{j_{1}}-{j_{2}}} (\bar x_{12})^{{\bar j}-{\bar
j_{1}}-{\bar j_{2}}}\[\Phi^{(j,\bar j)}\](x_{2},\bar x_{2}|z_{2}, \bar z_{2})\cr}\,} 
where
\eqn\defF{\[\Phi^{(j,\bar j)}\](x_{1},\bar
x_{1}|z_{1}, \bar z_{1})\,=\,{\cal R}^{j}_{j_1,j_2}(z_{12},x_{12})
\bar {\cal R}^{\bar j}_{\bar j_1,\bar
j_2}(\bar z_{12},\bar x_{12})\,\Phi^{(j,\bar j)}(x_{2},\bar x_{2}|z_{2},\bar z_{2})\,}
and $z_{12}=z_1-z_2$, $x_{12}=x_{1}-x_{2}$, etc. Operators ${\cal
R}^{j}_{j_1,j_2}$   are defined by series
\eqn\ser{ {\cal
R}^{j}_{j_1,j_2}(z_{12},x_{12})\,=\,\sum_{n=0}^{\infty}z^{n} \sum_{a_{i}=0,\pm
1} J^{a_{1}}_{-1}(x_{2}) J^{a_{2}}_{-1}(x_{2})\cdots
J^{a_{n}}_{-1}(x_{n})\,R_{n}(a_{1},..,a_{n})\,} 
where $R_{n}(a_{1},..,a_{n})$ are some differential operators
\eqn\defr{R_{n}(a_{1},..,a_{n})\,=\,\sum_{p=0}
r_{p}(a_{1},..,a_{n}|x_{12}) \(\frac{\d}{\d x_{2}}\)^{p}\,}
Operators $R_{n}$ are analogs of coefficients $\b_{m,n}^{l}$ and
$\b_{m,n}^{a,l}$ in \ \opeps\ . There is, of course, similar
expression for  $\bar {\cal R}^{\bar
j}_{\bar j_1,\bar j_2}$. In the following to simplify notations we
will often  omit all ``bar''
arguments and indices as we will work in the holomorphic sector.

Operators $R_{n}(a_{1},..,a_{n})$ can be obtained, as we already mentioned,
recursively from the requirement that both sides in \ \opeff\
transform in same way under $\wtl {su(2)_{k}}$. In particular to get
the recursion we apply $J_{n}^{a}(x)$ with $n=0$ and $n=1$ to both
sides of \ \opeff\ . Action with $J_{0}^{a}(x)$ relate different
$r_{p}$'s within same $R_{n}$ and action with $J_{1}^{a}(x)$ allow to
relate $r_{p}$'s for two successive $R_{n}$. The operator $R_{0}$ was
found in \zf . For $(j_1,j_2,j)=(2,\frac{k}{2},\frac{k}{2}-2)$ it
is identity operator
\eqn\ro{R_{0}\,=\,1\,}
It is possible to find the next operator $R_{1}(a)$. The only nonzero
coefficients $r_{p}(a)$ for the case $(j_1,j_2,j)=(2,\frac{k}{2},\frac{k}{2}-2)$ are 
\eqn\rn{\eqalign{&r_{0}(0)\,=\,\frac{2(k-6)}{k-4},\cr
&r_{0}(-)\,=\,\frac{4}{k-4},\cr
&r_{1}(-)\,=\,\frac{2}{k-4}\cr}\,}
Now, finally, we introduce functions
\eqn\deffunc{\eqalign{&f(x)\,=\,\sum_{p=-2}^{2} f_{p} x^{2+p}\cr
&g(x)\,=\,\sum_{p=-k/2}^{k/2} g_{p} x^{k/2+p}\cr
&h(x)\,=\,\sum_{p=-k/2+2}^{k/2-2} h_{p} x^{k/2-2+p}\cr}\,}
to define operators
\eqn\defoper{\eqalign{&\Phi^{(2)}(z)\,=\,\int d \mu^{(2)}
(x,\bar x) f(\bar x) \, \Phi^{(2)}(x,z)\,,\cr
&\Psi(z)\,=\,\int d \mu^{(\frac{k}{2})} (x,\bar x) g(\bar
x)\,\Phi^{(\frac{k} {2})}(x,z)\,,\cr
&\Phi^{(\frac{k}{2}-2)}(z)\,=\,\int d \mu^{(\frac{k}{2}-2)}
(x,\bar x) h(\bar x)\,\Phi^{(\frac{k}{2}-2)}(x,z)\cr}\,}
where $d \mu^{(j)}$ is a measure
\eqn\mes{d \mu^{(j)}(x,\bar x)=\frac{1}{\pi}\frac{d^2 x}{(1+x \bar x)^{2+2j}}\,}
We have arrived at  the problem: when the residue ${\rm
Res}\[\Phi^{(2)}(z_{1})\Psi(z_{2})\]_{z_1=z_2}$
is a total derivative 
\eqn\respr{{\rm
Res}\[\Phi^{(2)}(z_{1})\,\Psi(z_{2})\]_{z_1=z_2}\,=\,L_{-1}
\Phi^{(\frac{k}{2}-2)}(z)\,\,?\,}
The equation \ \respr\ is equivalent to the system of linear equations
\eqn\linsys{\eqalign{&\int d \mu^{(\frac{k}{2})} (x,\bar x) g(\bar x)
\((k-6)(a+1) p(x) x^{a}\,+\,p'(x) x^{a+1}\,-\,2 p(x) x^{a+1}
\frac{\d}{\d x}\)\,x^{\frac{k}{2}-2+m}\,=\cr
&=\,\frac{5}{2} \frac{k-4}{k+2} \int d \mu^{(\frac{k}{2}-2)}(x,\bar x) h(\bar
x) \(x^{a+1} \frac{\d}{\d x}\,-\,\half (k-4) (a+1) x^{a}\)
x^{\frac{k}{2}-2+m}\cr
&{\rm for}\,\,\,a\,=\{\,0,\pm
1\},\,\,\,\,\,\,m\,=\,\{-\frac{k}{2}+2,\cdots, \frac{k}{2}-2\} \cr}\,}
where 
\eqn\px{p(x)\,=\,f_{2}\,-\,f_{1} x\,+\,f_{0}
x^2\,-\,f_{-1} x^3\,+\,f_{-2} x^4\,}
This system has $3 k-9$ equations for $2 k-2$ variables
$g_{p}$, $h_{p}$, i.e. for $k \ge 8$ the system is overdetermined. We
have found solutions of this system for $f_{\pm 1}=0$ and
$k=6,7,8,\cdots,24$. Such a choice of $f_{p}$'s corresponds to the
perturbation \ \oex\ . From analysis of obtained solutions one can
draw

{\bf Conjecture.} (i) For any $k=4 n, 4 n+1,4 n+3 $ and $k \ge 11$ the
model  \ \per\ possesses
two additional IM which are certain linear combinations of currents
$\Psi_{m}$. (ii) For any $k=4 n + 2 \ge 14$ the model \ \per\
possesses three additional IM which are certain linear combinations of
currents $\Psi_{m}$.

Despite that the system \ \linsys\ has also solutions for
``resonance''  cases $k=6,7,8,10$,
we can not say anything about conservation of the
corresponding  charges in these cases. For first nonresonance cases
$k=11,12,13,14$  we  give solutions for $g_{p}$'s in Appendix D.

\newsec{Decay of particles.}

In section 3 we have constructed the solution describing decay process for 
the kink with mass $M_3$. The mass of that kink is always larger or
equal (if $A=B$) then sum of masses of other, lighter, two kinks  \
\maenq\ . It is natural to conjecture that while in the corresponding
quantum theory the heavy kinks are being excluded from ``in'' and ``out''
states they result in resonance poles in $S$-matrix. To justify this point
we calculated semiclassical limit of $S$-matrix for the
scattering $K_{01}^{(\ep_1)}(\th_{1})\,K_{13}^{(\ep_2)}(\th_{2})\,
\rightarrow\,K_{02}^{(\ep_2)}(\th_{2})\,K_{23}^{(\ep_1)}(\th_{1})$, \ \tgps\ 
\eqn\semclsol{\eqalign{S^{\rm (semicl)}_{K_{01} K_{13}}\,\equiv&\,\,{}_{\rm
out}<K_{02}^{(\ep_2)}({\th'}_{2})
K_{23}^{(\ep_{1})}({\th'}_{1})|K_{01}^{(\ep_{1})}(\th_{1})
K_{13}^{(\ep_2)}(\th_{2})>_{\rm in}\,=\cr
&\(\frac{16 \pi^{2}}{k}\)^2 \de ({\th'}_{1}-\th_{1}) \de
({\th'}_{2}-\th_{2})\,\exp\,\BL i \( {\cal A}_{K_{01}
K_{13}}\,-\, {\cal A}_{K_{01}}\,-
\,{\cal A}_{K_{13}} \) \BR \cr}\,}
where ${\cal A}_{g}$ is action functional $\cal A$ evaluated on the
corresponding kink solution \ \kink\ , \ \tgps\ . For any solution $g$
of \ \eqmo\ integrals on the left hand side in \ \semclsol\ can be
done  with the use of formula for a variation of $\cal A$
\eqn\vara{\de {\cal A}\[ g \]\,=\,-\frac{k}{8 \pi}\,{\rm Tr}\int d x d t\,
\d^{\mu}\((g^{-1} \de g) (g^{-1} \d_{\mu} g )\)\,}
One can obtain
\eqn\sems{\eqalign{S^{\rm (semicl)}_{K_{01} K_{13}}(\th_{12})\,=&\,
\(\frac{16 \pi^{2}}{k}\)^2 \de ({\th'}_{1}-\th_{1}) \de
({\th'}_{2}-\th_{2})\, \cdot\cr
&\exp\,\( -\frac{i \pi k}{4}-\frac{k}{2
\pi} \int_{0}^{\pi} d \zeta \ln \( \frac{1+e^{\th_{12}-\th_{0}-i
\zeta}}{e^{\th_{12}-\th_{0}}+e^{-i \zeta}}\) \)\cr} \,}
here $\th_{0}\,=\,\half \ln \((\al_1 \b_2)/(\al_2 \b_1)\)$,
$\th_{12}\,=\,\th_{1}-\th_{2}$.  From \
\sems\ follows that $S^{\rm (semicl)}_{K_{01} K_{13}}(\th)$ has a line of
essential singularities in the physical strip $0< Im \th < \pi$ at $Re
\th = \th_{0}$. We expect that this singular line corresponds to
series of resonance poles which become dense and are ``glued'' together as $k
\rightarrow \infty$.

We can also study scattering of ``basic'' particles
$\psi,\,\vth,\,\vphi$ which enter the action functional $\cal A$. From \ \aexp\
we read  their  masses 
\eqn\massbas{\eqalign{&m_{\psi}\,=\,2\, (k^{-1}\, \al_1 \b_1)^{1/2}\cr
&m_{\vth}\,=\,2\, (k^{-1}\, \al_2 \b_2)^{1/2}\cr
&m_{\vphi}\,=\,2\, (k^{-1}\, \al_3 \b_3)^{1/2}\cr}\,}
As in the kink case we obtain
\eqn\menq{m_{\vphi}\,\ge\,m_{\psi}\,+\,m_{\vth}}
This is an indication of instability of the $\vphi$ particle. Indeed,
to  the leading order we get
\eqn\decamp{{}_{\rm out} <\vth
(p_{3}) \psi(p_{2})|\vphi(p_{1})>_{in}\,=\,i\(\frac{16
\pi^{2}}{k}\)^{2} k^{-1} (\al_{1} \b_{2}\,-\,\al_{2}
\b_{1})\,\de^{(2)}(p_{1}\,-\,p_{2}\,-\,p_{3})}
In the rest frame of $\vphi$ particle $\psi$ particle propagate to
the right and $\vth$ propagate to the left.
Due to broken $P$-parity in the theory the amplitude for 
decay process with interchanged $\psi$ and $\vth$ is zero. 
It is possible to check that to the leading order  amplitudes for process
$\psi\,\psi \rightarrow  \vth\,\vth$ is  zero. This illustrates
integrability in  the tree approximation - nontrivial IMs prohibit
such kind of processes.  The amplitude for scattering $\psi$ and $\vth$ is
\eqn\ampdve{\eqalign{&S_{\vth \psi}(\th_{12})\,=\,{}_{\rm out}
<\psi({\th'}_{1}) \vth({\th'}_{2})|\vth(\th_{2})
\psi(\th_{1})>_{in}\,=\cr
 &\(\frac{16 \pi^{2}}{k}\)^{2} \(1\,+\,\frac{2 i}{k \sinh (\th_{12}\,\pm\,\th_{0})}\)
\de({\th'}_{1}\,-\,\th_{1}) \de({\th'}_{2}\,-\,\th_{2})\cr} \,}
here sign $\pm$ distinguish between $\psi$ particle propagating to the
left  or to the right in the center-of-mass frame.
Due to existence of higher IMs in the theory reflection part of
the amplitude is absent in \ \ampdve\ . The shift of
rapidity in \ \ampdve\ agrees with the one in semiclassical amplitude
for kinks \ \sems\ .  
\vfill
\eject

\newsec{Conclusions.}

In the paper we studied quantum integrability of $SU(2)_k$ WZW model
perturbed by multiplet of primary fields with isospin $j=2$. The model
contains four free parameters. The spectrum of the theory contains
kinks and their bound states. Kinks interpolate between different
vacua in the theory and has $Z_2$ charge. Some of kinks are unstable and we
obtained solution describing decay of the heaviest kink into two
lighter kinks. Quantum integrability was proven by explicit
construction of quantum IMs. It appears that there are two sets of
IMs. The first one - IMs belonging to universal enveloping algebra
$U\(\wtl {su(2)}_{k}\)$. Integrals of motion from this set admit semiclassical limit
$k \rightarrow \infty$ and there is recurrent formula for IMs in this
limit. The other set are integrals of motion built from chiral components
of primary field of isospin $j=k/2$ (conformal dimension
$\De_{\frac{k}{2}}=\frac{k}{4}$). Obviously these IMs do not admit
semiclassical limit. We also briefly discussed scattering amplitudes
for kinks and fundamental particles. We argued that scattering amplitudes for stable
kinks and particles possess resonance poles and calculated to the
leading order in $1/k$ decay amplitude for the heaviest particle. The
work on the construction of the exact kink S-matix of the theory is in
progress.

\newsec{Acknowledgments}

I would like to thank A.B.Zamolodchikov and S.L.Lukyanov for many
valuable discussions.

\newsec{Appendix A.}

Full expression for the two kink solution is given by

\eqn\twokink{\eqalign{&g_{11}\,=\,\frac{1-e^{2 \phi_{1}}\,+\,\De^{2} e^{2
\phi_{2}} - e^{2( \phi_{1}+\phi_{2})}}{1 + e^{2
\phi_{1}}\,+\,\De^{2} e^{2 \phi_{2}} + e^{2 (\phi_{1}+
\phi_{2})}}\,,\,\,\,\,\,g_{12}\,=\,\frac{2(-1)^{\ep_{1}} e^{\phi_{1}} (1-\De e^{2
\phi_{2}})}{1 + e^{2 \phi_{1}}\,+\,\De^{2} e^{2 \phi_{2}} + e^{2 (
\phi_{1}+\phi_{2})}}\cr
&g_{13}\,=\,\frac{2(-1)^{\ep_{1}+\ep_{2}} (1+\De) e^{\phi_{2} +\phi_{1}}}{1 + e^{2
\phi_{1}}\,+\, \De^{2} e^{2 \phi_{2}} + e^{2
(\phi_{1}+\phi_{2})}}\,,\,\,\,\,\,g_{21}\,=\,\frac{-2 (-1)^{\ep_{1}} e^{\phi_{1}} (1+\De e^{2
\phi_{2}})}{1 + e^{2 \phi_{1}}\,+\,\De^{2} e^{2 \phi_{2}} + e^{2 (
\phi_{1}+\phi_{2})}}\cr
&g_{22}\,=\,\frac{1-e^{2 \phi_{1}}\,-\,\De^{2}
e^{2 \phi_{2}} + e^{2 (\phi_{1}+\phi_{2})}}{1 + e^{2
\phi_{1}}\,+\,\De^{2} e^{2 \phi_{2}} + e^{2 (\phi_{1}+
\phi_{2})}}\,,\,\,\,\,\,g_{23}\,=\,\frac{-2 (-1)^{\ep_{2}} (
e^{2 \phi_{1}}-\De) e^{\phi_{2}}}{1 + e^{2
\phi_{1}}\,+\,\De^{2} e^{2 \phi_{2}} + e^{2 (\phi_{1}+
\phi_{2})}}\cr
&g_{31}\,=\,\frac{-2 (-1)^{\ep_{1}+\ep_{2}}  (1-\De) e^{\phi_{2} +\phi_{1}}}{1 + e^{2
\phi_{1}}\,+\, \De^{2} e^{2 \phi_{2}} + e^{2
(\phi_{1}+\phi_{2})}}\,,\,\,\,\,\,g_{32}\,=\,\frac{-2(-1)^{\ep_{2}} (\De+
e^{2 \phi_{1}}) e^{\phi_{2}}}{1 + e^{2
\phi_{1}}\,+\,\De^{2} e^{2 \phi_{2}} + e^{2 (\phi_{1}+
\phi_{2})}}\cr
&g_{33}\,=\,\frac{1+e^{2 \phi_{1}}\,-\,\De^{2}
e^{2 \phi_{2}} - e^{2 (\phi_{1}+\phi_{2})}}{1 + e^{2
\phi_{1}}\,+\,\De^{2} e^{2 \phi_{2}} + e^{2 (\phi_{1}+
\phi_{2})}}\cr}\,}

\newsec{Appendix B.}

Here we give expressions for first nontrivial IM's of spin $s=3$ in 
universal enveloping algebra $U\(\wtl {su(2)}_{k}\)$

\eqn\qintss{\eqalign{Q^{(3)}_{1}\,=\,\BL & 6 \,\eta_{-2}^{2} J^{-}_{-1} J^{-}_{-1} J^{-}_{-1}
J^{-}_{-1}\,+\,12 {\sqrt 6}\, \eta_{-2} \eta_{0} J^{-}_{-1} J^{-}_{-1}
J^{0}_{-1} J^{0}_{-1}\,+\cr
&20\, \eta_{-2} \eta_{2} J^{-}_{-1} J^{-}_{-1} J^{+}_{-1} J^{+}_{-1}\,+\,
16\, \eta_{-2} \eta_{2} J^{-}_{-1} J^{0}_{-1} J^{0}_{-1}
J^{+}_{-1}\,+\cr
&4(9\,\eta_{0}^{2}\,+\,2\,\eta_{-2} \eta_{2})\, 
J^{0}_{-1} J^{0}_{-1} J^{0}_{-1} J^{0}_{-1}\,+\,
12 {\sqrt 6}\, \eta_{0} \eta_{2} J^{0}_{-1} J^{0}_{-1} J^{+}_{-1}
J^{+}_{-1}\,+\cr
&6\, \eta_{2}^{2} J^{+}_{-1} J^{+}_{-1} J^{+}_{-1} J^{+}_{-1}\,+\,
12 {\sqrt 6}\,(8-k)\,\eta_{-2} \eta_{0} J^{-}_{-2} J^{-}_{-1}
J^{0}_{-1}\,+\cr
& 16 (k-10)\,\eta_{-2} \eta_{2}\,J^{-}_{-2} J^{0}_{-1}
J^{+}_{-1}\,+\,8 (k+2)\,\eta_{-2} \eta_{2}\,J^{0}_{-2} J^{-}_{-1}
J^{+}_{-1}\,+\cr 
& 6 {\sqrt 6}\,(12-k)\,\eta_{0} \eta_{2}\, J^{0}_{-2} J^{+}_{-1}
J^{+}_{-1}\,+\, 6 {\sqrt 6}\,(k-10)\, \eta_{-2}
\eta_{0}\,J^{-}_{-2} J^{-}_{-2}\,+\cr 
&(9 (k^2-12 k+16)\,\eta_{0}^{2}\,+\,2 (
k^2-8 k-4)\,\eta_{-2} \eta_{2})\,J^{0}_{-2} J^{0}_{-2}\,+\cr
&8 (k^2-13 k+24)\,\eta_{-2} \eta_{2}\, J^{-}_{-2} J^{+}_{-2}
 \BR \cdot 1\cr}\,}

\eqn\qintsss{\eqalign{Q^{(3)}_{2}\,=\,\BL & -6 \eta_{-2}^{2}\,J^{-}_{-1} J^{-}_{-1} J^{-}_{-1}
J^{-}_{-1}\,+\,12 {\sqrt 6}\,\eta_{-2} \eta_{0} J^{-}_{-1}
J^{-}_{-1} J^{-}_{-1} J^{+}_{-1}\,-\cr
&4 (6 \eta_{0}^{2}+5\eta_{-2}\eta_{2})\,J^{-}_{-1}
J^{-}_{-1} J^{+}_{-1} J^{+}_{-1}\,+\,8 (3 \eta_{0}^{2}-2
\eta_{-2}\eta_{2})\, J^{-}_{-1} J^{0}_{-1} J^{0}_{-1}
J^{+}_{-1}\,+\cr 
&4 (3 \eta_{0}^{2}-2\eta_{-2}\eta_{2})\,J^{0}_{-1}
J^{0}_{-1} J^{0}_{-1} J^{0}_{-1}\,+\, 
12 {\sqrt 6}\,\eta_{0} \eta_{2} J^{-}_{-1} J^{+}_{-1} J^{+}_{-1}
J^{+}_{-1}\,-\cr 
&6 \eta_{2}^{2}\,J^{+}_{-1} J^{+}_{-1} J^{+}_{-1} J^{+}_{-1}\,-\,
72 {\sqrt 6}\,\eta_{-2} \eta_{0}\,J^{-}_{-2} J^{-}_{-1}
J^{0}_{-1}\,+\cr
&(12 (k-16)\eta_{0}^{2}\,-\,2 (k+2) \eta_{-2} \eta_{2})
J^{0}_{-2} J^{-}_{-1} J^{+}_{-1}\,+\cr
&8 (k-10)\,(3 \eta_{0}^{2}\,-\,2\eta_{-2} \eta_{2})\,J^{-}_{-2}
J^{0}_{-1} J^{+}_{-1}\,+\,36 {\sqrt 6}\,\eta_{0}
\eta_{2}\,J^{0}_{-2} J^{+}_{-1} J^{+}_{-1}\,+\cr
&(6 (6+10 k-k^2)\,\eta_{0}^{2}\,-\,8 (k^2-13 k+24)\,\eta_{-2}
\eta_{2})\,J^{-}_{-2} J^{+}_{-2}\,+\cr
&(3 (k^2-8 k-40)\,\eta_{0}^{2}\,-\,2 (k^2-8
k-4)\,\eta_{-2} \eta_{2} ) \,J^{0}_{-2} J^{0}_{-2}\,+\cr 
&3 {\sqrt 6} (k-10) (k-2)\, \eta_{-2} \eta_{0}\,J^{-}_{-2}
J^{-}_{-2}\,+\cr
&3 {\sqrt 6} (k-10) (k-2)\, \eta_{0}
\eta_{2}\,J^{+}_{-2} J^{+}_{-2} \BR \cdot 1\cr}\,} 

\vfill
\eject

\newsec{Appendix C.}

In this appendix we prove that for integer $k$ current algebra $\wtl {su(2)}_{k}$ can
be extended by operators $\Psi_{m}$, \ \newcur\ .

In section $4.2$ we introduced generating functions $J(x,z)$ and
$\Phi^{(j,\bar j)}(x,\bar x|z,\bar z)$. Using Knizhnik-Zamolodchikov
equation and null-field $J_{-1}^{-}(x)\,\Phi^{(\frac{k}{2})}(x,z)$
one can derive \zf equation for four-point conformal block,
\eqn\conbl{G\,=\,<\Phi^{(\frac{k}{2})}(x_{1},z_{1})
\Phi^{(\frac{k}{2})}(x_{2},z_{2}) \Phi^{(\frac{k}{2})}(x_{3},z_{3})
\Phi^{(\frac{k}{2})}(x_{4},z_{4})>\,}
The solution of the equation  takes a simple form
\eqn\soleq{G^{12}_{34}(x,z)\,=\,\BL x_{32} x_{14}\, (x-z) \BR^{k} \BL z_{32} z_{14}\, z
(z-1) \BR^{-k/2}\,}
where $x=(x_{12} x_{34})/(x_{14} x_{32})$, $z=(z_{12} z_{34})/(z_{14}
z_{32})$. To check associativity of the algebra generated by
$\Phi^{(\frac{k}{2})}(x,z)$ we fuse $\Phi$'s in different ways and
compare results. For example, we can fuse pairs
$\Phi^{(\frac{k}{2})}(x_{1},z_{1}),\,\Phi^{(\frac{k}{2})}(x_{2},z_{2})$
and
$\Phi^{(\frac{k}{2})}(x_{3},z_{3}),\,\Phi^{(\frac{k}{2})}(x_{4},z_{4})$,
or
$\Phi^{(\frac{k}{2})}(x_{1},z_{1}),\,\Phi^{(\frac{k}{2})}(x_{3},z_{3})$
and
$\Phi^{(\frac{k}{2})}(x_{2},z_{2}),\,\Phi^{(\frac{k}{2})}(x_{4},z_{4})$.
To make functions $G^{12}_{34}(x,z)$ and $G^{13}_{24}(1-x,1-z)$ coincide
one should impose ``equal-time'' commutation relations
\eqn\comrel{\Phi^{(\frac{k}{2})}(x_{1},z_{1})
\Phi^{(\frac{k}{2})}(x_{2},z_{2})\,=\,e^{\frac{i \pi k}{2}}\Phi^{(\frac{k}{2})}(x_{2},z_{2})
\Phi^{(\frac{k}{2})}(x_{1},z_{1})\, ,\,\,\,\, \,}
With this prescription algebra generated by operators $\Psi_m$ become
associative.

\vfill
\eject

\newsec{Appendix D.}

Here solutions for functions $g(x)$, \ \deffunc\ are given;
$\al_{0,\pm 1}$, $\al$  and  $\b$ are free parameters.

\eqn\godi{\eqalign{&{\underline {k=11}}\cr 
&g_{\frac{11}{2}}\,=\,\frac{1}{3696} f_{0} (80 f_{0}^{2} f_{-2} f_{2}
- 7 f_{0}^{4} - 320 f_{-2}^2 f_{2}^2)\, \b \cr
&g_{\frac{9}{2}}\,=\,\frac{2}{3} f_{2}^5\, \al \cr
&g_{\frac{7}{2}}\,=\,\frac{5}{168} f_{-2} (16 f_{0}^{2} f_{-2} f_{2}
- f_{0}^{4} + 64 f_{-2}^2 f_{2}^2)\, \b \cr
&g_{\frac{5}{2}}\,=\,-5 f_{0} f_{2}^4\, \al \cr
&g_{\frac{3}{2}}\,=\,-\frac{5}{14} f_{0} f_{-2}^2 (f_{0}^2 + 24 f_{-2}
f_{2})\, \b \cr
&g_{\frac{1}{2}}\,=\,f_{2}^3 (5 f_{0} + 8 f_{_2} f_{2})\, \al \cr
&g_{-\frac{1}{2}}\,=\,f_{-2}^3 (5 f_{0} + 8 f_{_2} f_{2})\, \b \cr
&g_{-\frac{3}{2}}\,=\,-\frac{5}{14} f_{0} f_{2}^2 (f_{0}^2 + 24 f_{-2}
f_{2})\, \al \cr
&g_{-\frac{5}{2}}\,=\,-5 f_{0} f_{-2}^4\, \b \cr
&g_{-\frac{7}{2}}\,=\,\frac{5}{168} f_{2} (16 f_{0}^{2} f_{-2} f_{2}
- f_{0}^{4} + 64 f_{-2}^2 f_{2}^2)\, \al \cr
&g_{-\frac{9}{2}}\,=\,\frac{2}{3} f_{-2}^5\, \b \cr
&g_{-\frac{11}{2}}\,=\,\frac{1}{3696} f_{0} (80 f_{0}^{2} f_{-2} f_{2}
- 7 f_{0}^{4} - 320 f_{-2}^2 f_{2}^2)\, \al \cr}\,}

\eqn\gdve{\eqalign{&{\underline {k=12}}\cr &g_{6} ={1 \over 1485}
f_{2}^3\,(8\,f_{0}^2\, \beta + 27\,f_{-2}\,f_{2}\,\beta +
                     15\,f_{0}^2\,\alpha)\cr  & 
   g_{4} = {1 \over 90}f_{-2}\,f_{2}^3\,f_{0}\,(8\,\beta + 15\,\alpha)\cr  &
   g_{2} = f_{-2}^2\,f_{2}^3\,\beta\cr  & 
   g_{0} = f_{-2}^2\,f_{2}^2\,f_{0}\alpha\cr &
   g_{-2} = f_{-2}^3\,f_{2}^2\,\beta\cr  & 
   g_{-4} ={1 \over 90} f_{-2}^3\,f_{2}\,f_{0}\,(8\,\beta + 15\,\alpha)\cr &
   g_{-6} = {1 \over 1485} f_{-2}^3\,(8\,f_{0}^2\,\beta + 27\,f_{-2}\,f_{2}\,\beta +  
             15\,f_{0}^2\,\alpha)\cr
   &g_{-5}\,=\,g_{-3}\,=\,g_{-1}\,=\,g_{1}\,=\,g_{3}\,=\,g_{5}\,=\,0\cr}\,}

\eqn\gdve{\eqalign{&{\underline {k=13}}\cr
&g_{\frac{13}{2}}\,=\,\frac{1}{6864} (3 f_{0}^6 - 40 f_{0}^4 f_{-2}
f_{2} + 192 f_{0}^4 f_{-2}^2 f_{2}^2 - 512 f_{-2}^3 f_{2}^3)\, \b \cr
&g_{\frac{11}{2}}\,=\,-\frac{4}{11} f_{2}^6\, \al \cr
&g_{\frac{9}{2}}\,=\,\frac{1}{132} f_{0} f_{-2} ( f_{0}^4 - 16 f_{0}^2
f_{-2} f_{2} + 192 f_{-2}^2 f_{2}^2 )\, \b \cr
&g_{\frac{7}{2}}\,=\,4 f_{0}f_{2}^5\, \al \cr
&g_{\frac{5}{2}}\,=\,\frac{1}{12} f_{-2}^2 (f_{0}^4 - 48 f_{0}^2
f_{-2} f_{2} + 64 f_{-2}^2 f_{2}^2 )\, \b \cr
&g_{\frac{3}{2}}\,=\,-f_{2}^4 (7 f_{0}^2 + 8 f_{-2} f_{2})\, \al \cr
&g_{\frac{1}{2}}\,=\,2 f_{0} f_{-2}^3 (f_{0}^2 + 8 f_{-2} f_{2})\, \b \cr
&g_{-\frac{1}{2}}\,=\,2 f_{0} f_{2}^3 (f_{0}^2 + 8 f_{-2} f_{2})\, \al \cr
&g_{-\frac{3}{2}}\,=\,-f_{-2}^4 (7 f_{0}^2 + 8 f_{-2} f_{2})\, \b \cr
&g_{-\frac{5}{2}}\,=\,\frac{1}{12} f_{2}^2 (f_{0}^4 - 48 f_{0}^2
f_{-2} f_{2} + 64 f_{-2}^2 f_{2}^2 )\, \al  \cr
&g_{-\frac{7}{2}}\,=\,4 f_{0}f_{-2}^5\, \b \cr
&g_{-\frac{9}{2}}\,=\,\frac{1}{132} f_{0} f_{2} ( f_{0}^4 - 16 f_{0}^2
f_{-2} f_{2} + 192 f_{-2}^2 f_{2}^2 )\, \al \cr
&g_{-\frac{11}{2}}\,=\,-\frac{4}{11} f_{-2}^6\, \b \cr
&g_{-\frac{13}{2}}\,=\,\frac{1}{6864} (3 f_{0}^6 - 40 f_{0}^4 f_{-2}
f_{2} + 192 f_{0}^4 f_{-2}^2 f_{2}^2 - 512 f_{-2}^3 f_{2}^3)\, \al \cr}\,}
\vfill
\eject
\eqn\gche{\eqalign{&{\underline {k=14}}\cr &g_{7} = {1 \over
        15015}f_{2}^3\,(5\,f_{0}^3\,\alpha_{1} + 86\,f_{-2} f_{0} f_{2}\,\alpha_{1}+
                      35\,f_{0}^2 f_{2}\,\alpha_{-1} + 
                     77\,f_{-2} f_{2}^2\,\alpha_{-1})\cr  & 
    g_{6} = {5 \over 143}\,f_{2}^3\, \alpha_{0}\cr &  
    g_{5} = {1 \over 165}
            f_{-2} f_{2}^3\,(f_{0}^2\,\alpha_{1}+15\,f_{-2} f_{2}\,\alpha_{1}+
            7\,f_{0} f_{2}\,\alpha_{-1})\cr  &
    g_{4} = -{5 \over 11}\,f_{0} f_{2}^2\,\alpha_{0}\cr  & 
    g_{3} = {1 \over 15}f_{-2}^2 f_{2}^3\,(f_{0}\,\alpha_{1}+
             7\,f_{2}\,\alpha_{-1} )\cr  & 
    g_{2} = f_{2}\,(f_{0}^2 + f_{-2} f_{2})\,\alpha_{0}\cr  & 
    g_{1} = f_{-2}^3 f_{2}^3\,\alpha_{1}\cr  & 
    g_{0} = -{3 \over 7}\,f_{0}\,(f_{0}^2 + 6\,f_{-2} f_{2})\,\alpha_{0}\cr  &
    g_{-1} = f_{-2}^3\,f_{2}^3\,\alpha_{-1}\cr &
    g_{-2} = f_{-2}\,(f_{0}^2 + f_{-2} f_{2})\,\alpha_{0}\cr  &
    g_{-3} = {1 \over 15}f_{-2}^3 f_{2}^2\,(f_{0}\,\alpha_{-1} + 7\,f_{-2}\,\alpha_{1})\cr  & 
    g_{-4} = -{5 \over 11}\,f_{-2}^2 f_{0}\,\alpha_{0}\cr  &
    g_{-5} = {1 \over 165}f_{-2}^3 f_{2}\,(f_{0}^2\,\alpha_{-1} +
             15\,f_{-2} f_{2}\,\alpha_{-1} + 
             7\,f_{-2} f_{0}\,\alpha_{1})\cr  & 
    g_{-6} = {5 \over 143}\,f_{-2}^3\,\alpha_{0}\cr  & 
    g_{-7} = {1 \over 15015}f_{-2}^3\,(5\,f_{0}^3\,\alpha_{-1} +
             86\,f_{-2} f_{0}\,f_{2}\,\alpha_{-1}   
             +35\,f_{-2} f_{0}^2\,\alpha_{1} +
    77\,f_{-2}^2 f_{2}\,\alpha_{1})\cr}\,}

\vfill
\eject

\listrefs

\end